\documentclass[doublespacing]{elsart}

\usepackage{amssymb}
\usepackage{amsmath}
\usepackage{bm}
\usepackage{graphicx}

\begin{document}

\begin{frontmatter}



\title{High-spin states in $^{22}$Ne}


\author[spb]{S. Yu. Torilov}\ead{torilov@hotmail.com},
\author[abo]{M. Brenner},
\author[am]{V. Z. Goldberg},
\author[spb]{K. A. Gridnev},
\author[kri]{S. V. Khlebnikov},
\author[spb]{T. V. Korovitskaya},
\author[abo]{T. L\"onnroth},
\author[tu]{M. Mutterer},
\author[ki]{B. G. Novatski},
\author[abo]{J. Slotte},
\author[du]{Yu. G. Sobolev},
\author[jyfl]{W. H. Trzaska}, 
\author[kri]{G. P. Tyurin},
\author[spb]{L. I. Vinogradov}, 
\author[spb]{V. I. Zherebchevskiy}

\address[spb]{Department of Nuclear Physics, Saint-Petersburg State University, St. Petersburg, Russia}
\address[abo]{Department of Physics, Abo Akademi Universitet, Turky, Finland }
\address[jyfl]{Accelerator Laboratory, Department of Physics, University of Jyv\"askyl\"a, Finland}
\address[kri]{V.G.Khlopin Radium Institute, St.-Petersburg, Russia }
\address[tu]{Institut f\"ur Kernphysik, Technische Universitat, Darmstadt, Germany }
\address[ki]{Kurchatov Institute, Moscow, Russia}
\address[am]{Cyclotron Institute, Texas A\&M University, Texas, USA}
\address[du]{JINR, Dubna, Moscow, Russia }

\begin{abstract}
The structure of high-lying states in $^{22}$Ne has been studied using the $^{14}$C($^{12}$C,$\alpha$)$^{22}$Ne reaction at E($^{12}$C)= 44 MeV. The spins were determined by measuring double ($\alpha$,$\alpha$) angular correlations. Selective population of the 9$^-$ and 11$^-$ states at E$_x$=20.1 and 20.7 MeV, respectively, identifies those states as the 9$^-$ and 11$^-$ members of the first $K^{\pi}$ = 0$^-$ band, whose lower members were investigated by a method using inverse kinematics and a thick gas target. The spin and parity of four other new levels were determined to be 9$^-$ (21.5 MeV),12$^+$ (22.1 MeV),9$^-$ (25.0 MeV) and 8$^+$ (22.9 MeV). The two levels 9$^-$ and 12$^+$ may belong to the rotational doublets. 
\end{abstract}

\end{frontmatter}

\section{Introduction}
Interaction of cluster degrees of freedom with the extra neutrons of $sd$-shell region attracts more and 
more attention during the last few years due to the lack of the information about these nuclei.  The 
$^{22}$Ne nucleus is a promising candidate for this investigation, due to, as well known, a clear $\alpha$-
cluster structure of the self-conjugate $^{20}$Ne. The investigation of these states in $^{22}$Ne might be useful for the explanation and prediction of the some unusual properties of the system with extra neutrons, for instance, both of the new results for splitting of the band in $^{22}$Ne \cite{gold} and for the 
modern theories which predict stabilization of the Bose-condensation for nuclei with extra neutrons.  \\
A number of the $\alpha$-cluster states have perviously been found in $^{22}$Ne at excitation energies between 10-15~MeV \cite{gold}, which have been interpreted as members of unknown rotational band. From the other hand, some evidence was provided for the $^{18}$O+$\alpha$ breakup of  $^{22}$Ne from a number of states at a high excitation energy, in the range 12-24~MeV \cite{curt}, and for the $^{14}$C+$^{8}$Be breakup in the range 24-30~MeV \cite{fr}. These states may be linked to a  $^{18}$O+$\alpha$ and $^{14}$C+2$\alpha$ cluster structure in $^{22}$Ne. 
The aim of this research was to measure the spin and parity of some of these highly excited levels and possibly to investigate their nature and intrinsic configuration.

\section{Experiment}
\begin{figure}[h]
\begin{center}
\includegraphics[width=12cm,keepaspectratio,clip]{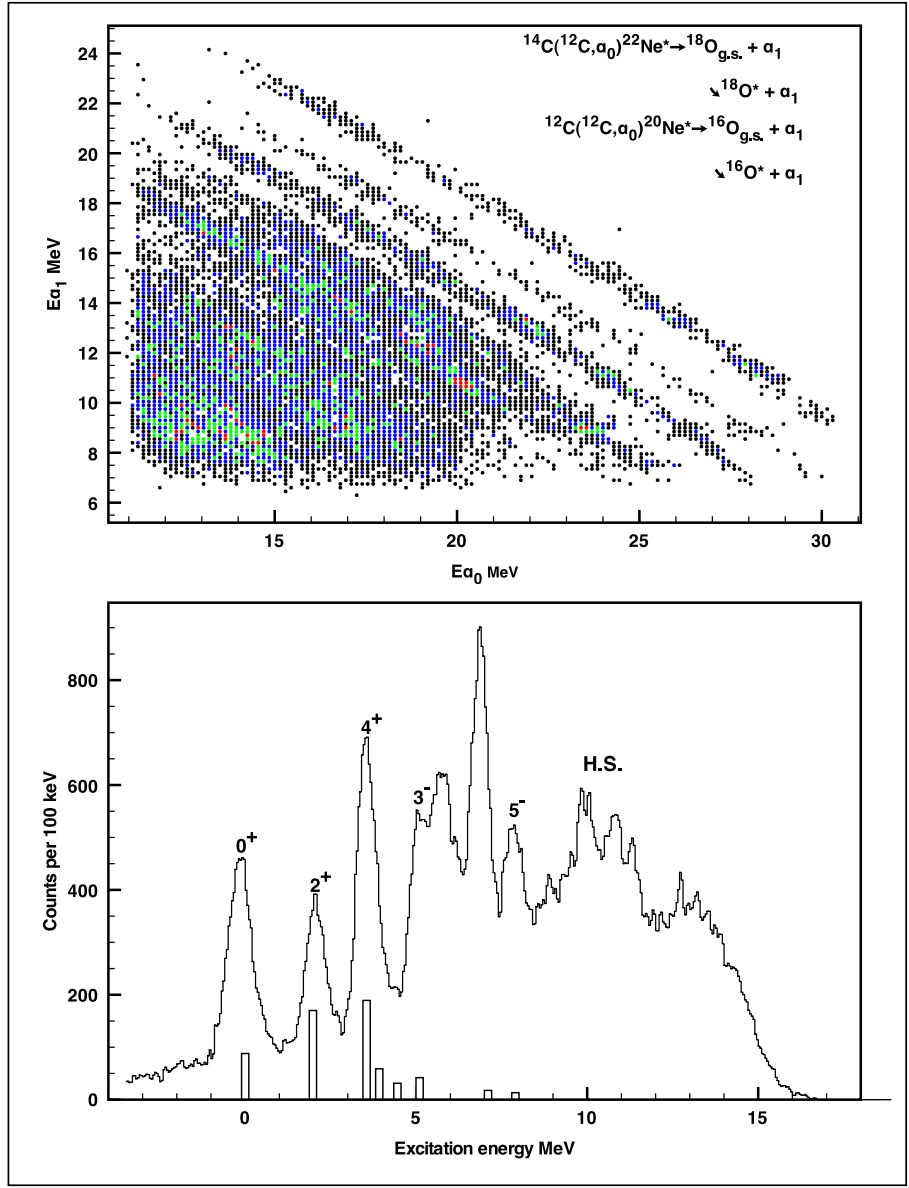}
\end{center}
\caption{Up: Two-dimensional plot of the energy of the decay alpha-particle ($\alpha_1$) which is detected in position sensitive detector versus the $\alpha_0$-particle energy.
Down: Spectrum of states of $^{18}$O excited by he $^{14}$C($^{12}$C,2$\alpha$). Three high spin states appearing in the continuum are indicated by H.S.. The empty bars are the Hauser-Feshabach calculation.} 
\label{fig:fig1}
\end{figure}

\begin{figure}[h]
\begin{center}
\includegraphics[width=12cm,keepaspectratio,clip]{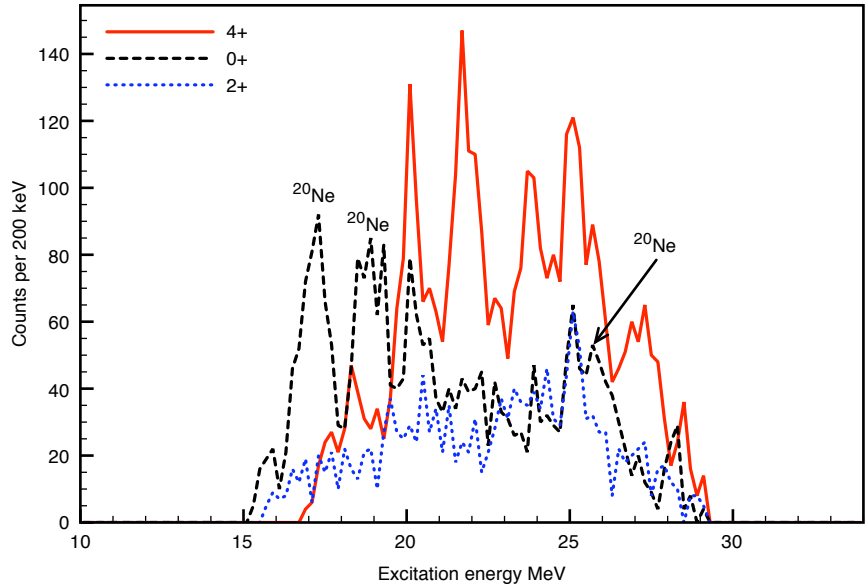}
\end{center}
\caption{The double coincidence spectra in which the $^{22}$Ne* states $\alpha$-decay to various final states in $^{18}$O (color online). }
\label{fig:fig2}
\end{figure}

\begin{figure}[h]
\begin{center}
\includegraphics[width=12cm,keepaspectratio,clip]{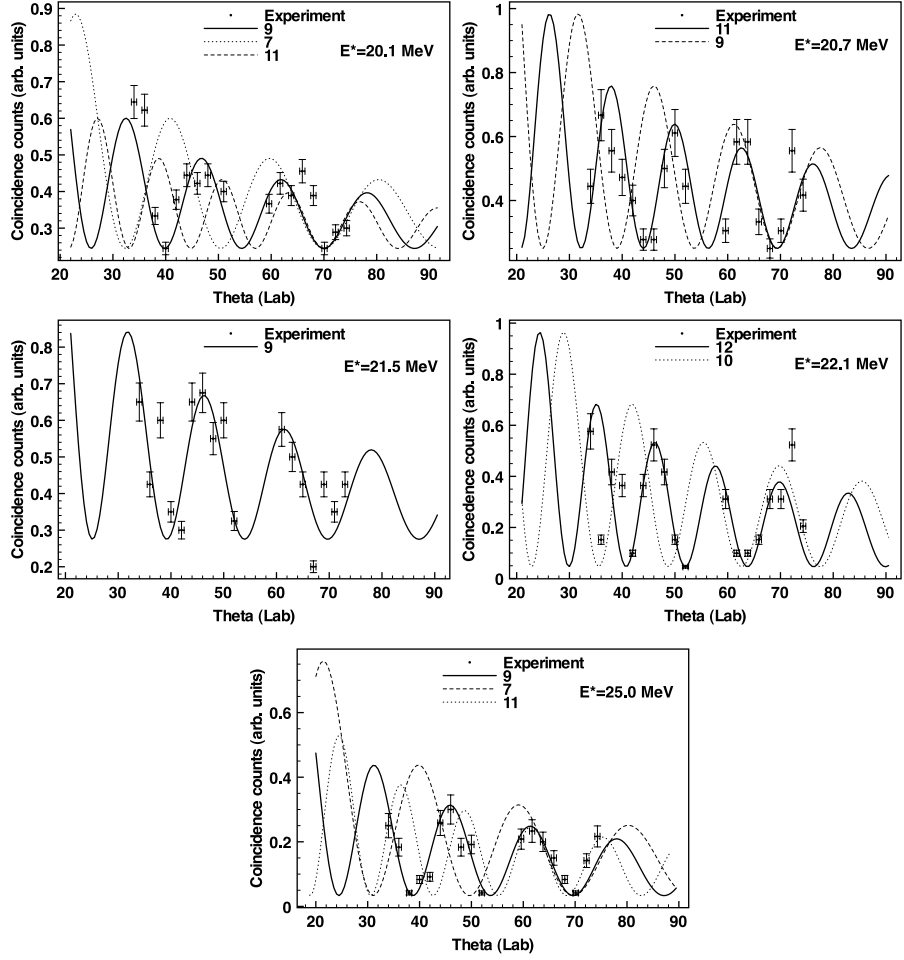}
\end{center}
\caption{Double angular correlations for the decay of the states in $^{22}$Ne at 20.1, 20.7, 21.5, 22.1 and 25.0 MeV to $^{18}$O$_{g.s.}$. The curves are fits using a $|P_J|^2+const.$ angular dependence.}
\label{fig:fig3}
\end{figure}

The reaction $^{12}$C+$^{14}$C$\to\alpha_0$+$^{22}$Ne*$\to\alpha_1$+$^{18}$O was used for this investigation because it was well known to preferentially populate high-spin states in $^{20}$Ne \cite{pan}, which subsequently could decay by $\alpha$-particle emission to a number of levels of $^{16}$O with known intrinsic configuration.
The experiment was carried out at the K-130 Cyclotron of the University of Jyv\"askyl\"a, Finland, with an 44 MeV $^{12}$C beam.  The target was a self-supported carbon foil of thickness 280 $\mu$m/cm$^2$ (80$\%$ of $^{14}$C). The beam was stopped in a Faraday cup and the accumulated charge, measured by a current integrator, was used for normalization in the measurements.
The primary $\alpha$-particle was detected using a pulse-shape detectors \cite{mu} at $\pm$3$^{\circ}$ behind a 15 $\mu$m platinum absorber used to stop the elastic $^{12}$C ions.  
Two groups detectors were placed for particle identification of reaction products in a scattering chamber of 1.5~m diameter. \\
The $\alpha$-particles from the decay of states in $^{22}$Ne were detected in a large area position-sensitive dE-E counter. It consists of a gas proportional counter used as dE detector combined with 10 silicon PIN diodes as a stop detectors. The total active area of the dE-E detector is 18 cm$^2$. The thickness of PIN – diodes depletion layer is 380 $\mu$m. The gas dE proportional counter of a single-wire type gives the information about the energy loss and the X-coordinate of the particles penetrating it. The counter was filled with Ar+10$\%$ CH$_4$ gas mixture (pressure – 250 Torr), continuous renewing of gas in the counter volume was provided.
The length of the detectors (100 mm) spanned 40$^{\circ}$ in the laboratory system (34$^{\circ}\ge\theta\ge$74$^{\circ}$). The angles at the center of each diode were measured to better than $\pm$0.5$^{\circ}$. \\
The $\alpha$-$\alpha$ double coincidence events were analyzed to generate the $\alpha_1$ angular distribution for decays to the ground state of $^{18}$O.
The decay channel was identified kinematically from a two-dimensional plot E$\alpha_1$ versus E$\alpha_0$ at each $\alpha_1$ angle (as defined by the position-sensitive of the $\alpha_1$ detector). An example of such a plot is shown in Fig.\ref{fig:fig1}. In fact the states of $^{18}$O can be reached in two ways via the coincidences: 1) $\alpha$-particles from the fusion-evaporation process (compound nucleus) are detected in coincidence and the total kinetic energy corrected for the missing recoil energy of $^{18}$O or 2) from the  transfer - breakup process. Spectrum for the excitation of $^{18}$O can be reconstructed from the coincident events and is shown in Fig.\ref{fig:fig1} (low part). The empty bars give the predictions of the Hauser-Feshbach calculation. The optical model parameters were taken from \cite{tam} and the level densities from \cite{huan}. Take into account that this is not a cross-section calculations, but for ration value.  We observe from Fig.\ref{fig:fig1} that Hauser-Feshbach mechanism reproduce the average behaviour of the cross section for the low-lying 0$^+$,2$^+$ and 4$^+$ states, but these calculations fail to reproduce the average behavior of the other excited states in $^{18}$O. From the other hand one can expect higher contribution of the direct processes for 0$^+$ and 4$^+$ states. Fig.\ref{fig:fig2} shows the double coincidence spectra for decay of $^{22}$Ne* to the ground state of  $^{18}$O, to the state at 1.98 MeV and to the state at 3.55 MeV. It is clear that the majority states decay primarily to the excited state 4$^+$ (3.55MeV). A striking feature of the results shown in the present work and in \cite{free} is the highly nonstatistical nature of the alpha decays from the resonances in $^{14}$C+$^{12}$C to excited states in $^{22}$Ne. \\
For the $\alpha$-$\alpha$ measurements described here the theoretical correlation reduces to the form :
\begin{equation}
W(\theta)= \frac{2l+1}{4\pi}|P_l(cos(\theta)|^2,\nonumber
\label{eq:hamiltonian}
\end{equation}
where $l$ is the spin of the level in $^{22}$Ne. It is assumed that the solid angle subtended by the 0$^{\circ}$ detector, which detects $\alpha_1$, is sufficiently small. The background, caused by $^{12}$C at forward angles, $\alpha$-particles, and particles from other possible reactions, was very small because of the small accidental – coincidence rate. 
Fig.\ref{fig:fig3} shows the double angular correlations and the resulting spin assignments for the states at 20.1 MeV (9$^-$), 20.7 MeV (11$^-$), 21.5 MeV (9$^-$), 22.1 MeV (12$^+$) and 25.0 MeV (9$^-$). \\
There is the good agreement between our results for the states in $^{20}$Ne and the known spin assignments for these states \cite{hi}

\section{Results}
\begin{figure}[h]
\begin{center}
\includegraphics[width=12cm,keepaspectratio,clip]{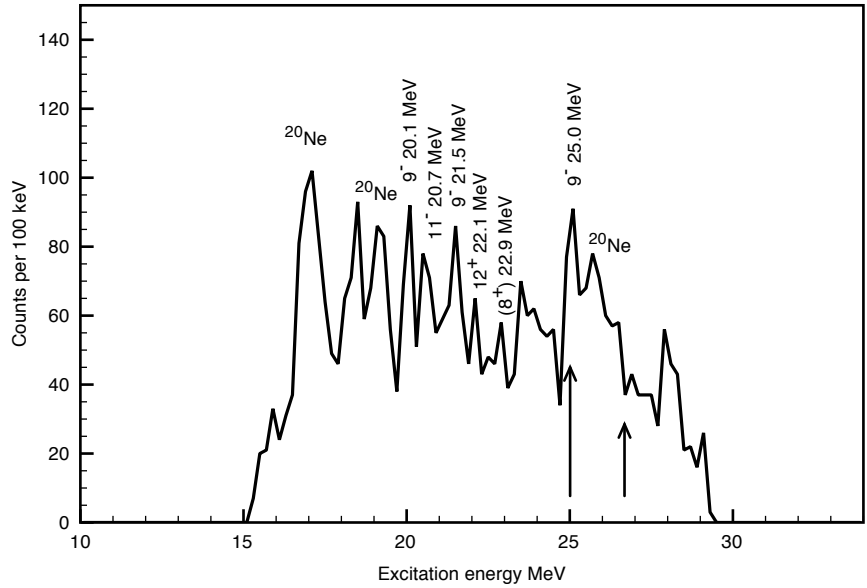}
\end{center}
\caption{The double coincidence spectrum, obtained at 44 MeV for the decays of the states in $^{22}$Ne to $^{18}$O. The states which were identified as corresponding to $^{14}$C+$^{8}$Be are indicated by arrows.}
\label{fig:fig4}
\end{figure}

\begin{figure}[h]
\begin{center}
\includegraphics[width=12cm,keepaspectratio,clip]{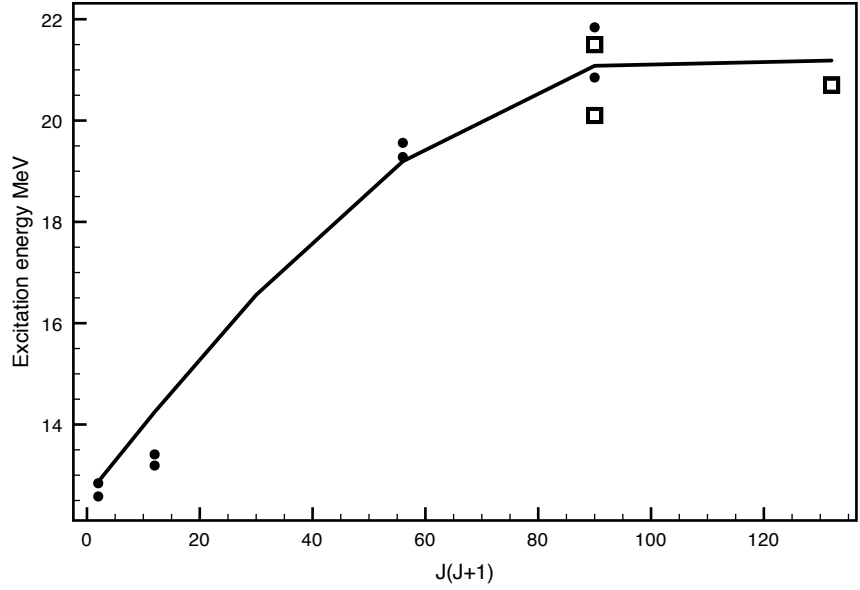}
\end{center}
\caption{Energies of negative parity states (dots - \cite{gold}, squares - the present work), together with the potential model calculation. $V_{WS}$ =-170 MeV, $a$=0.52 fm, $r_0$=1.25 fm, $r_C$=1.2 fm}
\label{fig:fig5}
\end{figure}

Figure \ref{fig:fig4} shows a sample of the $\alpha$-particles double coincidence spectra,obtained with 44-MeV incident beam, in which the $^{22}$Ne* states decay to the groung state in $^{18}$O. Buck $et$ $al.$ \cite{buck} had considerable success in explaining the properties of a number of states in light nuclei using a simple cluster model. In this model such states are considered as bound levels and resonances of a cluster-core system. The luster is treated as a single entity, with zero internal excitation, interacting with an inert core nucleus, and giving riseto states of relative motion with principal quantum number $N$ and orbital angular momenta $L$. These numbers are relates to the single-particle quantum numbers $n_j$ and $l_j$ of the nucleons in the cluster by 
\begin{equation}
2N+L= \sum_{j=1}^{n_c} (2n_j+l_j),\nonumber
\label{eq:hamiltonian}
\end{equation}
where $n_c$ is the number of cluster nucleons. To describe the rotational levels of $^{22}$Ne we need to place the cluster nucleons in the $sd$ shell or higher, i.e. $2N+L\ge8$. Goldberg $et$ $al.$ \cite{gold1} have done a calculation of the excited states in the $^{22}$Ne, using wave functions with five nodes for $L$=0 or 1. The number of nodes for other orbital momenta was calculated in accordance with the Wildermuth condition. In the present work the Woods-Saxon potential ($V_{WS}$=-170~MeV, $r_0$=1.25 fm, $a$=0.52) was used. Figure \ref{fig:fig5} presents the excitation energies for negative parity states together with the potential model calculations. As can be seen, the potential model with the same parameters describes the positions of the higher members of the negative parity band quite well. \\
Following the successful application of $\alpha$+$^{18}$O cluster models in $^{22}$Ne, attempts have been made to extend these models to include more complicated cluster configurations. Earlier evidence \cite{fr} indicated that two states in the $^{22}$Ne at 24.14 and 26.89 MeV decay by $^{8}$Be emission (they are indicated by arrows in Fig. \ref{fig:fig4}). As was shown in \cite{fr}, the $\alpha$-cluster states of  $^{20}$Ne and $^{22}$Ne should appear at similar excitation energy, measured with respect to the $\alpha$-decay threshold, and have similar properties. From this point of view there is a very good agreement between 9$^-$ of  $^{20}$Ne at 21.1 MeV (K$^{\pi}$=0$^-$ \cite{pan}) and 9$^-$ of  $^{22}$Ne at 25.0 MeV (present work).

\section{Summary}
The $^{14}$C($^{12}$C,$\alpha_0$)$^{22}$Ne reaction has been studied at 44~MeV. The coincident detection of the $\alpha_0$ and $\alpha_1$ breakup from the decay of excited states in $^{22}$Ne has allowed the spins of the decaying states to be studied. The data provide evidence that a molecular rotational bands are populated in the reaction. The measured rotational parameters indicates that the structure has a large moment of inertia and is consistent with a quasimolecular configuration. 

\section*{Acknowledgements}
The participation of two of the coauthors (S.T and V.Z.) in the experiments was granted by the Magnus Ehrnrooth Foundation trough the Finnish Society of Science and Letters



\newpage

\newpage

\newpage

\newpage

\newpage

\end{document}